\documentclass[prl,superscriptaddress,showpacs,twocolumn]{revtex4}

\usepackage[pdftex]{graphicx}

\usepackage{graphicx,amsfonts}

\usepackage{epsfig,amsmath, hyperref}

\usepackage{verbatim}

\usepackage[english]{babel}

\newcommand{\Z}{\mathbb{Z}}
\newcommand{\R}{\mathbb{R}}

\renewcommand{\P}{\mathbb{P}}
\newcommand{\E}{\operatorname{E}}
\newcommand{\envP}{\E_Q}
\newcommand{\Pl}{\mathrm{\textbf{P}}}

\newcommand{\Var}{\operatorname{Var}}

\newcommand{\Ai}{\operatorname{Airy}}
\newcommand{\Aifunc}{\operatorname{Ai}}
\newcommand{\Bi}{\operatorname{Bi}}

\begin{document}

\title{The Intermediate Disorder Regime for Directed Polymers in Dimension $1+1$}

\author{Tom Alberts}
\email{alberts@math.toronto.edu}
\affiliation{Department of Mathematics, University of Toronto, Toronto, ON, Canada}

\author{Kostya Khanin}
\email{khanin@math.toronto.edu}
\affiliation{Department of Mathematics, University of Toronto, Toronto, ON, Canada}

\author{Jeremy Quastel}
\email{quastel@math.toronto.edu}
\affiliation{Department of Mathematics, University of Toronto, Toronto, ON, Canada}

\date{\today}

\begin{abstract}
We introduce a new disorder regime for directed polymers with one space and one time dimension that is accessed by scaling the inverse temperature parameter $\beta$ with the length of the polymer $n$. We scale $\beta_n := \beta n^{-\alpha}$ for $\alpha \geq 0$. This scaling sits in between the usual weak disorder ($\beta = 0$) and strong disorder regimes ($\beta > 0$). The fluctuation exponents $\zeta$ for the polymer endpoint and $\chi$ for the free energy depend on $\alpha$ in this regime, with $\alpha = 0$ corresponding to the usual polymer exponents $\zeta = 2/3, \chi = 1/3$ and $\alpha \geq 1/4$ corresponding to the simple random walk exponents $\zeta = 1/2, \chi = 0$. For $\alpha \in (0, 1/4)$ the exponents interpolate linearly between these two extremes. At $\alpha = 1/4$ we exactly identify the limiting distribution of the free energy and the end point of the polymer.
\end{abstract}

\pacs{05.40.-a, 02.50.-r, 75.10.Nr}

\maketitle

\section{Introduction}

Directed polymers in disordered media is a model of a variety of physical phenomena. It describes vortex lines in superconductors \cite{Nelson:superconductor}, domain walls \cite{HuseHenley}, roughness of crack interfaces \cite{HHR}, Burgers turbulence \cite{Burgers:airy} and the KPZ equation \cite{KPZ:growth_interface}. It provides a simple model in which the more difficult predictions of spin glasses can be tested and is related \cite{HuseHenley} to the phase boundary in the 2D Ising model. Starting with \cite{HuseHenley, forster_nelson_stephen:fluid}, they have been the subject of intensive study \cite{HalpinHealy_Zhang, HuseHenleyFisher:prl} over the last twenty-five years. In the setting of the $d$-dimensional integer lattice, the polymer measure is a random probability measure on the set of paths of $d$-dimensional nearest neighbour lattice walks. The randomness of the polymer measure is inherited from a random environment which manifests itself as an i.i.d. collection of random variables placed on the sites of $\Z_+ \times \Z^d$. Given a fixed environment $\omega : \Z_+ \times \Z^d \to \R$, the energy of an $n$-step nearest neighbour walk $s$ is
\begin{eqnarray*}
& H_n^{\omega}(s) = \sum_{i=1}^n \omega(i, s_i). &
\end{eqnarray*}
The polymer measure on such walks is then defined in the usual Gibbsian way by
\begin{align*}
\Pl_{n, \beta}^{\omega}(s) =  e^{-\beta H_n^{\omega}(s)} \, \Pl(s)/Z_{n, \beta}^{\omega},
\end{align*}
where $\beta > 0$ is the inverse temperature, $\Pl$ is the uniform measure on $n$-step walks, and $Z_n^{\omega}(\beta; x)$ the point-to-point partition function
\begin{align*}
Z_n^{\omega}(\beta; x) = \E_\Pl \big[ e^{-\beta H_n^{\omega}(s)} \delta_0(s(n) = x) \big].
\end{align*}
The partition function $Z_n^{\omega}(\beta)$ is the sum of the point-to-point versions over $x$. The random environment is a probability measure $Q$ on the space of environments $\Omega = \{ \omega : \Z_+ \times \Z^d \to \R \}$. Let $Q$ be a product measure so that the variables $\omega(i,z)$ are independent and identically distributed, and assume they have mean zero and variance one and
$\lambda(\beta) := \log \envP e^{-\beta \omega} < \infty$.

The goal is to study the behaviour of the model as $\beta$ and $d$ vary. At $\beta = 0$ the polymer measure is the simple random walk, hence the walk is entropy dominated and exhibits diffusive behaviour. For $\beta$ large the polymer measure concentrates on paths with low energy and the diffusive behaviour is no longer guaranteed. A more precisely defined separation between these two regimes is given in terms of the quenched and annealed free energies:
\begin{align*}
F_q(\beta) \! = \! \lim_{n \uparrow \infty} \! \frac{E_Q \log Z_n^{\omega}(\beta)}{n} \! \leq \! \lim_{n \uparrow \infty} \! \frac{\log E_Q Z_n^{\omega}(\beta)}{n} = \lambda(\beta).
\end{align*}
Weak disorder corresponds to equality and manifests itself through diffusive behavior. In the opposite, strong disorder case the system exhibits localization. The transition occurs at a critical $\beta_c$. For $d = 1$ and $2$, $\beta_c = 0$, while $0 < \beta_c \leq \infty$ for $d \geq 3$ \cite{ImSpen:diffusion}. We only consider $d=1$.

The behaviour of the polymer is described in terms of fluctuation exponents for the polymer endpoint and the free energy. For $1+1$-dimensional polymers there are long-standing predictions: For the polymer endpoint $s(n)^2 \sim n^{2\zeta}$ with $\zeta = 2/3$ for all $\beta$; for the free energy fluctuation exponent $\Var_Q (\log Z_{n}^{\omega}(\beta)) \sim n^{2\chi}$, with $\chi = 1/3$. Observe that $\zeta = 2/3$ and $\chi = 1/3$ satisfy the relation $\chi = 2 \zeta - 1$, which is expected to hold whenever there is localization.

For the free energy the limiting distribution of the fluctuations is the Tracy-Widom distribution \cite{TracyWidom:level_spacing} corresponding to the Gaussian Unitary Ensemble of random matrices, that is $c(\beta)(\log Z_n^{\omega}(\beta) - F_q(\beta)n )/n^{1/3} \to F_{GUE}$. In fact, a stronger statement holds for the point-to-point partition functions. After proper normalization the collection of partition functions parameterized by the renormalized target point $x$ converges to a universal stationary process called Airy$_2$ \cite{PrahoferSpohn:airy_process} which has $F_{GUE}$ as its one-dimensional distributions. Precisely
\begin{align}\label{AiConvergence}
\frac{\log Z_{n}^{\omega}(\beta; n^{2/3}s) - F_q(\beta)n}{c(\beta)n^{1/3}} + a(\beta)s^2 \to \Ai_2(s),
\end{align}
where $a(\beta)$ is a positive constant which compensates for a convex dependence of the free energy on the target point, and $Z_n^{\omega}(\beta; x)$ is the point-to-point partition function with endpoint at $x$.

We introduce the localization length exponent $\nu$:
\begin{align*}
\Var_{\Pl_{n, \beta}^{\omega}}(s(n)) = E_{\Pl_{n, \beta}^{\omega}} \left[ \left( s(n) - E_{\Pl_{n, \beta}^{\omega}}[s(n)] \right)^2 \right] \sim n^{2\nu}.
\end{align*}
In the weak disorder (diffusive) case $\nu = 1/2$. In the strong disorder case $\nu = 0$ for typical environments, i.e. the polymer is localized. This does not contradict the known fact that the variance averaged over environments is of order $n$ \cite{HuseFisher:paths_potential}. For very few environments of total probability on the order of $n^{-1/3}$, $\Var_{\Pl_{n, \beta}^{\omega}}(s(n))$ is on the order of $n^{4/3}$. This explains the apparent discrepancy.

\section{Exponents for Intermediate Disorder}

The central focus of this paper is to describe these exponents and the limiting distributions in what we have termed an intermediate disorder regime, accessed by scaling $\beta$ as we increase the polymer length,
\begin{align}\label{beta_scaling}
\beta_n := \beta n^{-\alpha}, \quad \alpha \geq 0.
\end{align}
The $\alpha = 0$ case is the much studied polymer measure, with predictions given on the previous page. For different values of $\alpha \geq 0$ the ``$\alpha$-polymers'' behave in ways that are quantifiably different from each other. In this sense the intermediate disorder regime is actually an entire family of regimes with $\alpha$ as an indexing parameter. The value $\alpha = 1/4$ is critical and is also the most interesting; it is described soon. For $\alpha > 1/4$ the polymer is diffusive and behaves as a simple random walk, with $\sqrt{n}$ fluctuations of the path, i.e. $\zeta = 1/2$, $\nu = 1/2$. The polymer, rescaled by $\sqrt{n}$, converges to Brownian motion with diffusivity constant $1$ for almost all environments. Also $\chi = 0$. For $\alpha \leq 1/4$ the exponents $\zeta, \chi,$ and $\nu$ depend on $\alpha$ as
\begin{align*}
\zeta(\alpha) = {\small\frac{2}{3}}(1-\alpha), \,\,
\chi(\alpha) = {\small\frac{1}{3}}(1-4\alpha), \,\,
\nu(\alpha) = 2\alpha.
\end{align*}
The leading behaviour of $\log Z_n^{\omega}(\beta n^{-\alpha})$ is of order $n^{1-2\alpha}$.

The values of $\zeta(\alpha)$ and $\chi(\alpha)$ still satisfy $\chi(\alpha) = 2 \zeta(\alpha) - 1$. They linearly interpolate between the KPZ \cite{KPZ:growth_interface, Kardar:growth} scalings of $\zeta = 2/3, \chi = 1/3, \nu = 0$ at $\alpha = 0$ and the diffusive exponents $\zeta = 1/2, \chi = 0, \nu = 1/2$ at $\alpha = 1/4$. We now show how the interpolation can be derived from the values at $\alpha = 0$ together with the Airy process asymptotics of the point-to-point partition function. Let $E_n = \E_{\Pl_{n, \beta_n}^{\omega}}[ s(n) ]$ and assume for the moment that $n^{2 \nu}$ is of smaller order than $E_n$. This assumption means that most of the contribution to the fluctuations of the free energy comes from paths with an endpoint in a neighbourhood of $E_n$. The logarithm of the number of such paths is $-E_n^2/n$ (by the normal approximation to simple random walk) and from \eqref{AiConvergence} each path contributes energy on the order of
$n^{1/3-\alpha} \left[ \Ai_2(E_n/n^{2/3}) - \Ai_2(0) \right]$ to $\log Z_{n}^{\omega}(\beta; E_n)$, so that the log of the point-to-point partition function at $E_n$ is
\begin{align}\label{asymptotic_free_energy}
-\frac{E_n^2}{n} + n^{1/3 - \alpha} \left[ \Ai_2(E_n/n^{2/3}) - \Ai_2(0) \right].
\end{align}
The polymer endpoint favors the value of $E_n$ which maximizes \eqref{asymptotic_free_energy}. We obtain
\begin{align*}
E_n^2 \sim n^{4/3 - \alpha} \left[ \Ai_2(E_n/n^{2/3}) - \Ai_2(0) \right].
\end{align*}
It follows that $E_n$ is of a smaller order than $n^{2/3}$, and since on small distances the $\Ai_2$ process is similar to the Wiener process we have $\Ai_2(E_n/n^{2/3}) - \Ai_2(0) \sim \sqrt{|E_n|/n^{2/3}}$. Hence
\begin{align*}
E_n^2 \sim n^{1-\alpha} \sqrt{|E_n|},
\end{align*}
which gives $|E_n| \sim n^{2/3(1-\alpha)}$. Substituting this back into \eqref{asymptotic_free_energy} implies that the fluctuations of the free energy are on the order of $n^{1/3(1-4\alpha)}$, which gives the formula for $\chi(\alpha)$. The localization length exponent is determined by the condition that the difference between logarithms of the point-to-point partition function with endpoints separated by the localization length $n^{\nu}$ should be of order $1$. For distances $|x_1 - x_2| \sim n^{\nu(\alpha)}$ this difference is of the order of
\begin{align*}
n^{1/3 - \alpha} \left[ \Ai_2(x_1/n^{2/3}) - \Ai_2(x_2/n^{2/3}) \right],
\end{align*}
which is of order $\sqrt{x_1 - x_2} n^{-\alpha}$. This is of order $1$ only if $\nu(\alpha) = 2\alpha$. We immediately see that for $\alpha < 1/4$ the polymer is localized ($\nu < \zeta$).

Using the Wiener process approximation the limiting distribution for the pair
\begin{align}\label{energy_endpoint_pair}
&\left( \frac{\log Z_n^{\omega}(\beta_n; x n^{\zeta(\alpha)}) - \log Z_n^{\omega}(\beta_n; 0)}{c(\alpha, \beta) n^{\chi(\alpha)}}, \frac{E_{P_{n, \beta_n}^{\omega}}[s(n)]}{C(\alpha, \beta) n^{\zeta(\alpha)}} \right)
\end{align}
is the joint distribution of the maximum value $M$ of $W(t) - t^2$ and the point $t_M$ where the maximum is achieved. Here $W(t)$ is a $2$-sided Brownian motion. This probability distribution was calculated exactly in \cite{Groene}. The joint density of $(t_M, M)$ at $(t,a)$ is $g(|t|) h_a(|t|) \psi_a(0)$, where $g$ has Fourier transform
$
\hat{g}(\lambda) = \int \! e^{i \lambda s} g(s) \, ds = \xi \Aifunc(i \xi \lambda)^{-1},
$
$h_a$ has Laplace transform
\begin{align*}
\hat{h}_a(\lambda) = \! \int_{0}^{\infty} \!\!\!\! e^{-\lambda s} h_a(s) \, ds = \Aifunc(\rho a + \xi) \Aifunc(\xi)^{-1},
\end{align*}
and $\psi_a(x)$ has Fourier transform
\begin{align*}
\hat{\psi}_a(\lambda) = \pi \xi \left(\Aifunc(i \xi \lambda) \! \Bi(i \xi \lambda + \rho a) - \Bi(i \xi \lambda) \! \Aifunc(i \xi \lambda + \rho a)  \right)
\end{align*}
with $\xi = 2^{-1/3}, \rho = 2^{2/3}$. Here $\Aifunc$ and $\Bi$ are the Airy functions \cite{AbSteg}.

For $\alpha = 1/4$ the localization length is on the same order as the displacement of its endpoint $(\nu = \zeta)$. The polymer is not localized anymore and the latter argument breaks down. However the exponents are still correct. The scaling behavior in this critical case will be discussed in the next section.

We verify the formulas for $\zeta(\alpha)$ and $\chi(\alpha)$ by numerical simulations. We fix $\beta = 1$ and let $\alpha$ vary from $0$ to $.275$ in increments of $.025$. Exponents are computed for $500$ independent Gaussian environments and polymers of length $n = 5000$. See the figures for results.

\begin{figure}
\begin{center}
\includegraphics[width=8.5cm, height=4.25cm]{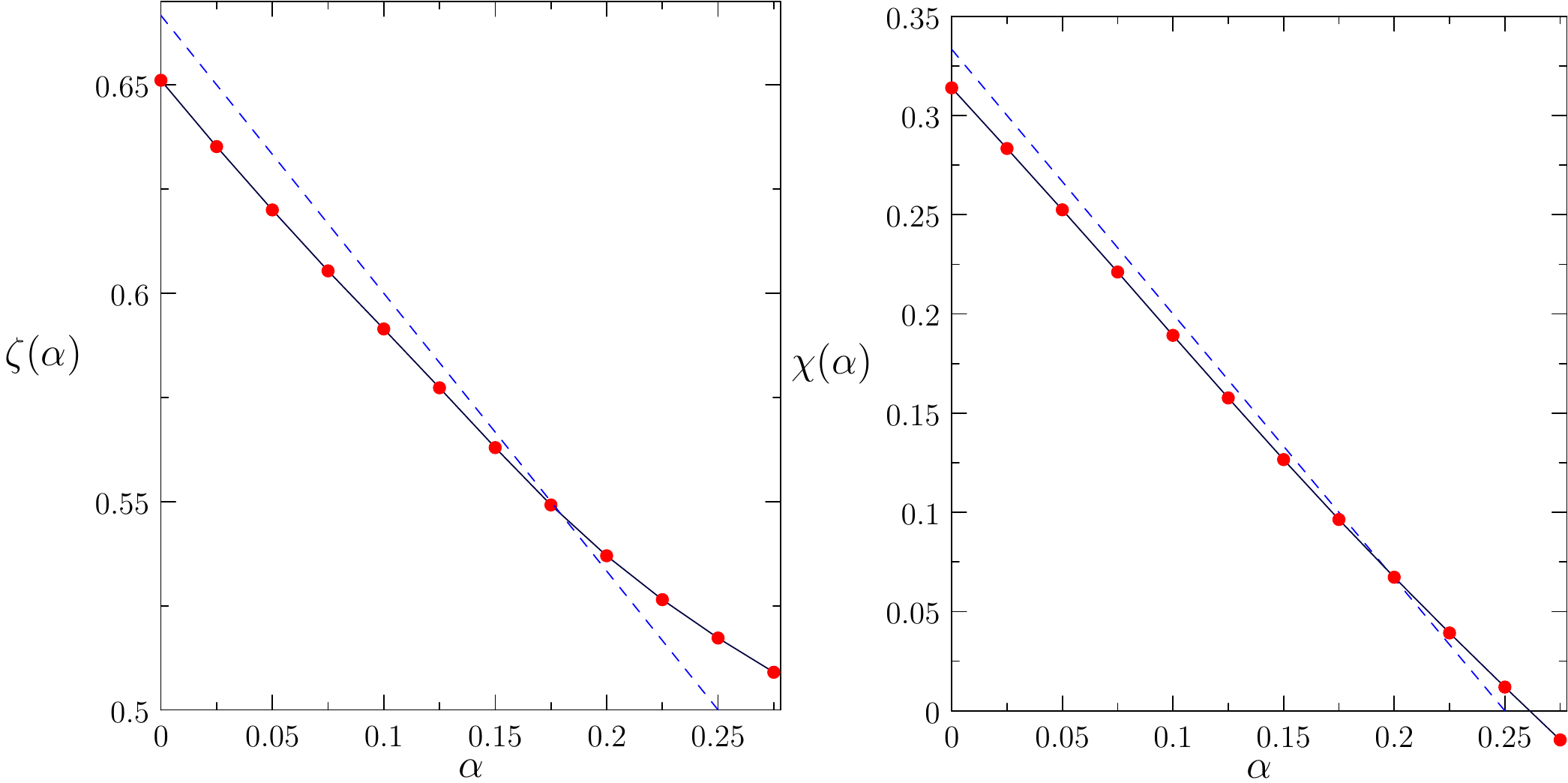}
\end{center}
\end{figure}

\section{Critical Regime}

In the $\alpha = 1/4$ case the limiting distribution of the polymer endpoint is more involved but still computable. From $\chi(1/4) = 0$ we do not require any normalization. Consider the modified partition function
\begin{align*}
\overline{Z}_{n}^{\omega}(\beta_n; x) = \E_\Pl \left[ \prod_{i=1}^n (1 + \beta n^{-\frac{1}{4}} \omega(i, s_i)) \delta(s(n) = x) \right].
\end{align*}
Expanding the product and then applying the expectation with respect to the random walk yields $\overline{Z}_n^{\omega}(\beta_n; x) = \sum_{k=0}^n \beta^k n^{-k/4} J_k^n(x)$, where
\begin{align*}
J_k^n(x) = \! \sum p_x^n(i_k, x_k) \prod_{j=1}^k \omega(i_j, x_j) p(i_j - i_{j-1}, x_j - x_{j-1}).
\end{align*}
Here $p(i,y) = \Pl(s(i) = y)$ for random walks starting at zero, $p_x^n(i_k, x_k) = p(n-i_k, x-x_k)$, and the sum is over ordered $1 \leq i_1 < \ldots < i_k \leq n$ with $i_0 = 0$, and free $x_j$, $1 \leq j \leq k$, with $x_0 = 0$. We compute that for fixed $k$
\begin{align*}
n^{-k/4} J_k^n(x\sqrt{n}) / p(n, x \sqrt{n}) \to 2^k I_k(x) / \varrho(1,x),
\end{align*}
where
\begin{align*}
I_k(x) = \! \int \! \varrho_x(t_k,x_k) \prod_{i=1}^k W(dt_i, dx_i) \varrho(t_i - t_{i-1}, x_i - x_{i-1}).
\end{align*}
Here $W(t,x)$ is a Gaussian white noise on $\R_+ \times \R$ with $\E[W(t,x) W(s,y)] = \delta(t-s) \delta(x-y)$, $\varrho(x,t) = \exp \{-x^2/2t \}/\sqrt{2\pi t}$, $\varrho_x(t_k, x_k) = \varrho(1-t_k, x-x_k)$, and the integration is over $\{0 = t_0 < t_1 < \ldots t_k \leq 1 \} \times \R^k$ with $x_0 = 0$. The $n^{-k/4}$ term is necessary to keep the variance of each term bounded, and the $2^k$ terms come from the local limit theorem for simple random walk. In the case $k=1$ the convergence is easily seen by a Fourier transform computation. We have $\lambda(i \beta) = -\frac{1}{2} \beta^2 (1 + o(1))$ as $\beta \to 0$ from the mean zero, variance one assumption on $\omega$, so
\begin{align*}
& \log  \E_Q  \left[ \exp \left \{ it n^{-\frac{1}{4}} J_1^n(x \sqrt{n})/p(n, x\sqrt{n}) \right \} \right] \\
&= -\frac{t^2 \beta^2}{2}(1+o(1)) n^{-1/2} \sum_{i=1}^n \sum_y q_n^x(i,y)^2 \to -\frac{2t^2 \beta^2}{\sqrt{\pi}}.
\end{align*}
Here $q_n^x(i,y) = p(i,y)p(n-i, x\sqrt{n} - y)/p(n, x\sqrt{n})$. Therefore the limiting distribution is normal with mean zero and variance $4 \beta^2/\sqrt{\pi}$, and it is easily checked that $2I_1(x)/\varrho(1,x)$ has the same distribution. For the $k > 1$ terms the convergence is handled by the general theory of U-statistics \cite{Janson:GHS}.

From these computations we conclude that $\overline{Z}_n^{\omega}(\beta_n; x\sqrt{n})/p(n, x\sqrt{n})$ converges to the process $\mathcal{Z}_{\beta}(x) := \sum_{k \geq 0} (2\beta)^k I_k(x)/\varrho(1,x)$. The same convergence also holds replacing $\overline{Z}$ by $Z$. Define the process $A_{\beta}(x) = \log \mathcal{Z}_{\beta}(x) - \log \varrho(1,x)$. It is very important that this process is universal, namely it appears as a limit for polymer models in the critical regime with an extremely general distribution for the random potentials $\omega$. It is also easy to see that the statistics of the density of the scaled polymer endpoint $s(n)/\sqrt{n}$ is the same as the statistics of the random density $C_{\beta} \exp \{ A_{\beta}(x) - x^2/2 \} \, dx$.

Another important property of $A_{\beta}(x)$ is that it interpolates between a Gaussian process ($\beta = 0$) and the Airy$_2$ process ($\beta = \infty$). To see this crossover one has to properly rescale the process with $\beta$ to normalize the variance of its one-point distribution and its two-point correlation function at a fixed distance. The interpolation property follows from \cite{ACQ} and \cite{SpohnSasa:KPZ, SpohnSasa:prl}, where an exact formula for the distribution of $A_{\beta}(x)$ was calculated (see also \cite{calabrese}). Using the Tracy-Widom formula for asymmetric simple exclusion \cite{TracyWidom:asep_integral_formulas, TracyWidom:asep_step_condition}, they obtain:
\begin{align*}
\P & \left \{ A_{\beta}(x) \geq s \right \} = \!\! \int \!  e^{-e^{-r}} \!\! f \! \left( \! s + 2\beta/3 - \log \sqrt{32 \pi \beta^4} - r \! \right) dr
\end{align*}
where,
\begin{align*}
f(r) = \kappa^{-1} \det(I-K_{\sigma_T})\mathrm{tr}\left((I-K_{\sigma_T})^{-1}\rm{P}_{\Ai}\right)
\end{align*}
with $\kappa = 2 \beta^{4/3}$, $\rm{P}_{\Ai}(x,y) = \Aifunc(x) \Aifunc(y)$, and
\begin{align*}
K_\sigma(x,y) = \int \sigma(t) & \Aifunc(x+t) \Aifunc(y+t) \, dt \\
& + \pi \kappa^{-1} G_{\frac{x-y}{2}} \left( \frac{x+y}{2} \right).
\end{align*}
Here $\sigma(t) = (1 - e^{- \kappa t})^{-1} - (\kappa t)^{-1}$ and
\begin{align*}
G_q(x) = \frac{2}{\pi^{3/2}} \int_0^{\infty} \frac{\sin \left(x \eta + \eta^3/12 - q^2/\eta + \pi/4 \right)}{\sqrt{\eta}} \, d\eta.
\end{align*}
In particular it follows that $\P \{ A_{\beta}(x) \leq 2\beta^{4/3} s \} \to F_{GUE}(s)$. It also follows that $A_{\beta}(x)$ is stationary in $x$ as its distribution is independent of $x$. Finally, one can show that the process $A_{\beta}(x)$ is continuous and locally Brownian, i.e. $A_{\beta}(x+\delta) - A_{\beta}(x)$ is of order $\sqrt{\delta}$.

\section{Exact Calculation of Exponents}

The limiting distribution is also characterized \cite{ACQ, SpohnSasa:KPZ} by observing that $\mathcal{Z}_{\beta}$ is the Wick exponential of a stochastic integral. Let $B$ be a one-dimensional Brownian motion that is independent of the white noise $W(t,x)$. Let $\E_{0,x}$ denote expectation over Brownian paths started at zero and ending at $x$ at time $1$. Then
\begin{eqnarray*}
\mathcal{Z}_{\beta}(x) = \E_{0,x} \left[ :\! \exp \! : \! \left \{ 2\beta \int_0^1 W(s, B(s)) \, ds \right \} \right].
\end{eqnarray*}
The exact density of the polymer endpoint $\mathcal{Z}_{\beta}(x) \varrho(1,x)$ in the critical regime is part of the critical polymer measure on paths. Define the continuum partition function
\begin{eqnarray}\label{alternate_representation}
Z_T^{W}(\beta) = \E_0 \left[ :\! \exp \! : \! \left \{ 2\beta \int_0^T W(s, B(s)) \, ds \right \} \right].
\end{eqnarray}
The polymer measure $\Pl_{n, \beta_n}^{\omega}$ in the critical $\alpha = 1/4$ case scales to a probability measure on paths of length $T$ given by
\begin{align*}
\mathrm{d} \Pl_{T, \beta}^W(B) =  :\! \exp \! : \! \left \{ 2\beta \int_0^T W(s, B(s)) \right \} \, \frac{\mathrm{d} \Pl(B)}{Z_T^W(\beta)},
\end{align*}
where $\Pl$ stands for the usual Wiener measure. Write $\E_{T, \beta}^W$ for expectation. \eqref{alternate_representation} provides another derivation of the fluctuation exponents $\zeta(\alpha)$ and $\chi(\alpha)$. For $\beta > 0$ let
\begin{align*}
\tilde{B}(s) := \beta^2 B(\beta^{-4} s), \,\, \tilde{W}(t, x) := \beta^{-3} W(\beta^{-4}t, \beta^{-2} x).
\end{align*}
Scaling properties imply that $\tilde{B}$ is a standard Brownian motion and $\tilde{W}$ is a standard space-time white noise, so via the substitution $u = \beta^4 s$ we have the relation
$
\beta \int_0^T W(s, B(s)) \, ds  =  \int_0^{\beta^4 T} \tilde{W}(u, \tilde{B}(u)) \, du
$
and so $Z_{\beta, T}^{W} = Z_{1, \beta^4 T}^{\tilde{W}}$. From
$
\Var_W(Z_{\beta, T}^{W}) \sim T^{2/3}
$
we get
\begin{align*}
\Var_W(Z_{T^{-\alpha}, T}^{W}) = \Var_W(Z_{1, T^{1-4\alpha}}^{W}) \sim T^{\frac{2}{3}(1-4\alpha)},
\end{align*}
which agrees with $\chi(\alpha)$. For $\zeta(\alpha)$ we have $\E_{T, \beta}^W \left[ B(T)^2 \right] \sim T^{4/3}$ as $T \to \infty$ for almost all $W$ and fixed $\beta$. Since $B(T)^2 = \beta^{-2} \tilde{B}(\beta^4 T)^2$ we have
\begin{align*}
\E_{T, \beta}^W \big[ B(T)^2 \big] = \beta^{-4} \E_{\beta^4 T, 1}^{\tilde{W}} \big [ \tilde{B}(\beta^4 T)^2 \big ].
\end{align*}
Therefore
\begin{align*}
\E_{T, T^{-\alpha}}^W \big[ B(T)^2 \big] &= T^{4\alpha} \E_{T^{1-4\alpha}, 1}^{\tilde{W}} \big[ \tilde{B}(T^{1-4\alpha})^2 \big] \sim T^{\frac{4}{3}(1-\alpha)}.
\end{align*}

\section{Conclusion}

We have identified a new disorder regime for directed polymers in $d=1+1$ and computed the corresponding wandering exponent $\zeta$ and free energy fluctuation exponent $\chi$, in addition to a localization length exponent $\nu$. These exponents vary linearly with the parameter $\alpha$ between the KPZ scaling exponents at $\alpha = 0$ and the simple random walk exponents at $\alpha = 1/4$. At $\alpha = 1/4$ there is a phase transition from intermediate disorder to weak disorder. For $\alpha > 0$ we are able to identify the joint limiting distribution of the free energy and the polymer endpoint, and give an explicit asymptotic formula for the distribution of the point-to-point free energy at $\alpha = 1/4$.

The research of all authors is supported by the Natural Sciences and Engineering Research Council of Canada.

\end{document}